\documentclass[acmsmall]{acmart}

\usepackage{enumitem}
\usepackage{amsmath}
\usepackage{pifont}
\usepackage[frozencache,cachedir=_minted-main]{minted}
\usepackage{array}
\usepackage{booktabs}
\usepackage{multirow}
\usepackage{algorithm}
\usepackage{algorithmicx}
\usepackage{algpseudocode}
\usepackage{subcaption}
\usepackage[capitalize]{cleveref}

\newcommand{\name}[1]{Documentary}
\newcommand{\code}[1]{\texttt{\small#1}}
\newcommand{\tick}{\ding{51}}
\newcommand{\cross}{\ding{55}}
\setminted{
  frame=lines,
  linenos,
  fontsize=\footnotesize,
  framesep=2mm,
  numbersep=5pt,
  xleftmargin=13pt,
  breaklines=true,
  breaksymbol=,
}

\setcopyright{cc}
\setcctype{by}
\acmDOI{10.1145/3808211}
\acmYear{2026}
\acmJournal{PACMSE}
\acmVolume{3}
\acmNumber{FSE}
\acmArticle{FSE204}
\acmMonth{7}
\acmSubmissionID{fse26mainb-p2869-p}
\received{2025-09-12}
\received[accepted]{2026-03-24}

\begin{document}

\title{Natural Language-Focused Software Engineering via Code-Documentation Equivalence}

\author{Aryaz Eghbali}
\orcid{0000-0001-9763-8147}
\affiliation{%
  \institution{CISPA Helmholtz Center for Information Security}
  \city{Stuttgart}
  \country{Germany}
}
\email{aryaz.egh@gmail.com}

\author{Zhongxin Liu}
\orcid{0000-0002-1981-1626}
\affiliation{%
  \institution{Zhejiang University}
  \city{Hangzhou}
  \country{China}
}
\email{liu_zx@zju.edu.cn}

\author{Michael Pradel}
\orcid{0000-0003-1623-498X}
\affiliation{%
  \institution{CISPA Helmholtz Center for Information Security}
  \city{Stuttgart}
  \country{Germany}
}
\email{michael@binaervarianz.de}

\begin{abstract}
  Source code documentation is an integral part of software development and maintenance, as it helps in understanding the code and facilitates communication among developers.
  However, existing documentation is often incomplete, outdated, or inaccurate, which can lead to misunderstandings and errors.
  In the era of large language models (LLMs), which are being extensively used for software engineering tasks, the quality of documentation becomes even more critical, as documentation provides important context for the models.
  In this paper, we introduce the notion of \emph{documentation-to-code equivalence}, a novel property that captures whether documentation accurately and completely describes the code it documents.
  We present a novel approach, called \emph{\name{}}, to automatically generate equivalent documentation for a given code snippet.
  Our evaluation shows that \name{} can generate equivalent documentation for 53.4\% of the evaluated function-level code snippets.
  To show the benefits of documentation-to-code equivalence, we describe and evaluate two software engineering tasks: code understanding and code editing.
  Our results show that documentation-to-code equivalence allows an LLM to predict the output of a function with 12.8--24.5\% higher accuracy, when compared to human-written documentation and documentation generated by a baseline approach.
  Furthermore, human developers consider documentation generated by \name{} to be more useful for understanding and editing code than the original human-written documentation.
\end{abstract}



\maketitle

\section{Introduction}
Source code, written in formal programming languages, is often difficult to read and understand, especially for complex codebases or when the code is written by someone else.
Hence, most programming languages support natural language documentation in the form of comments and docstrings, allowing developers to describe code intent more clearly.
Such documentation has been shown to be beneficial in improving code comprehension and facilitating collaboration among developers~\cite{Hu2020,Lin2023}.
However, time constraints and the continuous evolution of code can lead to documentation that is incomplete, outdated, or inaccurate~\cite{DeSouza2005_study,Hu2020,Lin2023,Liu2021_JIT}.
Such documentation, or simply no documentation at all, can hinder the understanding of the code and delay the development process.

\begin{figure}
  \centering
  \begin{minted}[firstnumber=739]{python}
...
  def pbr(self) -> None:
      "Pretty print has a line break"
      if self.p_p == 0:
          self.p_p = 1

  def p(self) -> None:
      "Set pretty print to 1 or 2 lines"
      self.p_p = 1 if self.single_line_break else 2

  def soft_br(self) -> None:
      "Soft breaks"
      self.pbr()
      self.br_toggle = "  "

  def o(self, data: str, puredata: bool = False, force: Union[bool, str] = False) -> None:
      """
      Deal with indentation and whitespace
      """
      if self.abbr_data is not None:
          self.abbr_data += data
...
  \end{minted}
  \caption{Part of the source code from the \code{Alir3z4/html2text} repository.}
  \label{fig:running_example_original}
\end{figure}

As an example of suboptimal documentation, consider the function \code{soft\_br} from the repository \code{Alir3z4/html2text}, shown in \cref{fig:running_example_original}.
This function is only four lines long, yet a developer or a large language model (LLM) unfamiliar with the codebase would struggle to understand its purpose and how it works without spending time or tokens reading other parts of the codebase.
For example, the function calls the \code{pbr} function, which itself sets some state in the object, and the effect of this state change is not clear without reading the code of other functions that use this state.
The \code{soft\_br} function also modifies the \code{br\_toggle} attribute, which is used in other parts of the code to determine how to format line breaks, and this is equally unclear without reading those functions.
Since this is only a short excerpt from a file exceeding 1,000 lines, identifying the relevant context is time-consuming.

Prior work on improving documentation focuses either on summarizing code~\cite{Allamanis2016,Wan2018,Zhang2020,Ahmad2020,Wang2021,Liu2021,LeClair2021} or on ensuring that documentation remains consistent with the code during software evolution~\cite{Liu2020_JIT,Panthaplackel2020a,Panthaplackel2021,Liu2021_JIT}.
While useful, these approaches do not ensure that the documentation is complete and accurate.
A summary typically omits details, and hence, does not ensure completeness.
Consistent documentation only guarantees soundness---it does not contradict the code---but makes no claim about completeness.
To illustrate the limitations of summarization and consistency, again consider the function \code{soft\_br} in \cref{fig:running_example_original}.
The docstring of the function is consistent with the code, and the function is short enough to not require any summarization.
However, the main limitation is that the docstring does not provide a complete description of the function's behavior, particularly in terms of its side effects and interactions with other parts of the code.
Another limitation of summarization-focused and consistency-focused approaches is that they do not provide a clear and objective way to evaluate the quality of the documentation.
The reason is that one side of the evaluation of the documentation-code relation is in natural language, and therefore unstructured and vague.
Prior work in these areas~\cite{Liu2021,Liu2020_JIT,Liu2021_JIT} has focused on evaluating metrics that measure how close the documentation is to human-written documentation.
However, such metrics of textual similarity may not accurately capture the semantic similarity of documentation and are limited because human-written documentation is often incomplete.
Yet, because of the lack of an objective measure for documentation quality, comparison to human-written documentation has become the de facto standard.

\begin{table}
  \caption{Comparison of documentation-code properties.}
  \label{tab:property_comparison}
  \centering
  \begin{tabular}{lccl}
    \toprule
    Doc-code property & Sound & Complete & Evaluation of property \\
    \midrule
    Consistency~\cite{Fluri2009} & \tick{} & \cross{} & Code vs. natural language \\
    Summarization~\cite{Allamanis2016} & \tick{} & \cross{} & Code vs. natural language \\
    Equivalence (this work) & \tick{} & \tick{} & Code vs. code \\
    \bottomrule
  \end{tabular}
\end{table}


    
    

To address the limitations of prior work, we introduce a novel property called \emph{documentation-to-code equivalence}, which captures whether a piece of documentation accurately and completely describes the corresponding code.
Informally, this property means that the documentation contains enough information to re-implement the same functionality as the code that it documents (see \cref{sec:definition} for a formal definition).
This is evaluated by checking whether an LLM can generate code that is equivalent to the original code, based on the documentation and the surrounding context.

\Cref{tab:property_comparison} summarizes the differences between the existing properties of documentation-code relations and our proposed property of documentation-to-code equivalence.
A key difference is that, in addition to ensuring the soundness of the documentation, i.e., that the documentation does not contradict the code, documentation-to-code equivalence also ensures the completeness of the documentation, i.e., that the documentation fully describes the code.
Moreover, it enables the evaluation of the documentation-code relation by evaluating the equivalence of two code snippets, which is a more well-defined problem than textual similarity and can be evaluated automatically.
\Cref{fig:running_example_generated_docstring} shows an equivalent docstring for the same \code{soft\_br} function from \cref{fig:running_example_original}, which provides a comprehensive explanation of the function's purpose, behavior, and interactions with other components of the codebase.

While documentation-to-code equivalence is a strong property, existing documentation often does not satisfy it, as illustrated by the example above.
To address this problem, we present a novel approach, called \emph{\name{}}, to automatically generate equivalent documentation for a given code snippet.
The key idea of \name{} is to iteratively refine the documentation by generating code from the documentation and comparing it to the original code, until the generated code is equivalent to the original code.
To determine the equivalence of two code snippets, we use an LLM as a judge~\cite{Ahmed2025,Bavaresco2025}, which allows our approach to be language- and platform-agnostic.

We envision documentation-to-code equivalence to be a useful property for various software engineering tasks, performed either by humans or LLMs.
In this paper, we explore two such tasks: code understanding and code editing.
For code understanding, we consider the task of predicting the output of a function for a given input, based on the function signature and its documentation.
For code editing, we consider the task of modifying code based on a natural language description of the desired change.

Our evaluation applies our ideas to function-level docstrings of real-world Python projects.
The results show that \name{} effectively generates equivalent docstrings for 53.4\% of the functions in our dataset, as determined by running the tests.
Regarding the code understanding task, we find that equivalent docstrings allow an LLM to predict the output of a function with 12.8--24.5\% higher accuracy, when compared to human-written documentation and documentation generated by a baseline approach.
For the code editing task, we find that equivalent docstrings help developers understand the code better than the human-written documentation.
These results demonstrate the benefits of documentation-to-code equivalence in improving the performance of LLMs and human developers in software engineering tasks.

With the recent surge in the use of LLMs for software engineering tasks, documentation is now also used by LLM-based tools and agents.
This means that the documentation should be useful not only for human developers, but also for LLMs.
Furthermore, the medium of communication between developers and LLMs is to a large extent natural language.
Therefore, we see documentation-to-code equivalence as an important step toward \emph{natural language-focused software engineering}, i.e., performing software engineering tasks partially or fully based on natural language documentation as a significant source of information.

In summary, this paper makes the following contributions:
\begin{itemize}
    \item A novel property (documentation-to-code equivalence) that describes a strong semantic relationship between code and its documentation.
    \item An iterative approach, called \name{}, to automatically generate equivalent documentation for a given code snippet.
    \item Empirical evidence of the benefits of equivalent documentation in two common software engineering tasks, code understanding and code editing.
\end{itemize}

\section{Approach}
We first define the novel concept of \emph{documentation-to-code equivalence} as a property of a piece of documentation relative to a code snippet (\cref{sec:definition}).
Then, we present our \name{} approach to automatically generate equivalent documentation for a given code snippet (\cref{sec:approach details}).

\subsection{Definition of Documentation-to-Code Equivalence}
\label{sec:definition}

Suppose we have a piece of code $c$, a piece of documentation $d$ that is intended to document $c$, and some surrounding context $\mathit{ctx}$ that provides additional information about the code, such as other functions, classes, or modules in the same file or project.
Furthermore, let $L$ be a large language model (LLM) that can generate code from natural language prompts.
The \emph{documentation-to-code equivalence} property is defined as follows:
The documentation $d$ is equivalent to the code piece $c$, with respect to an LLM $L$ and the surrounding context $\mathit{ctx}$, if $L$ can generate $c$ from $(d, ctx)$.
If this property holds, we say that $d$ is an \emph{equivalent documentation} for $c$.

The above formulation is general and can be instantiated with different kinds of code, documentation, context, and LLMs.
In the remainder of this paper, we focus on function-level docstrings as the documentation, function bodies as the code, and the whole source code file as the surrounding context.

\subsection{Generating Equivalent Documentation}
\label{sec:approach details}

\begin{figure}
\centering
\includegraphics[width=0.8\textwidth]{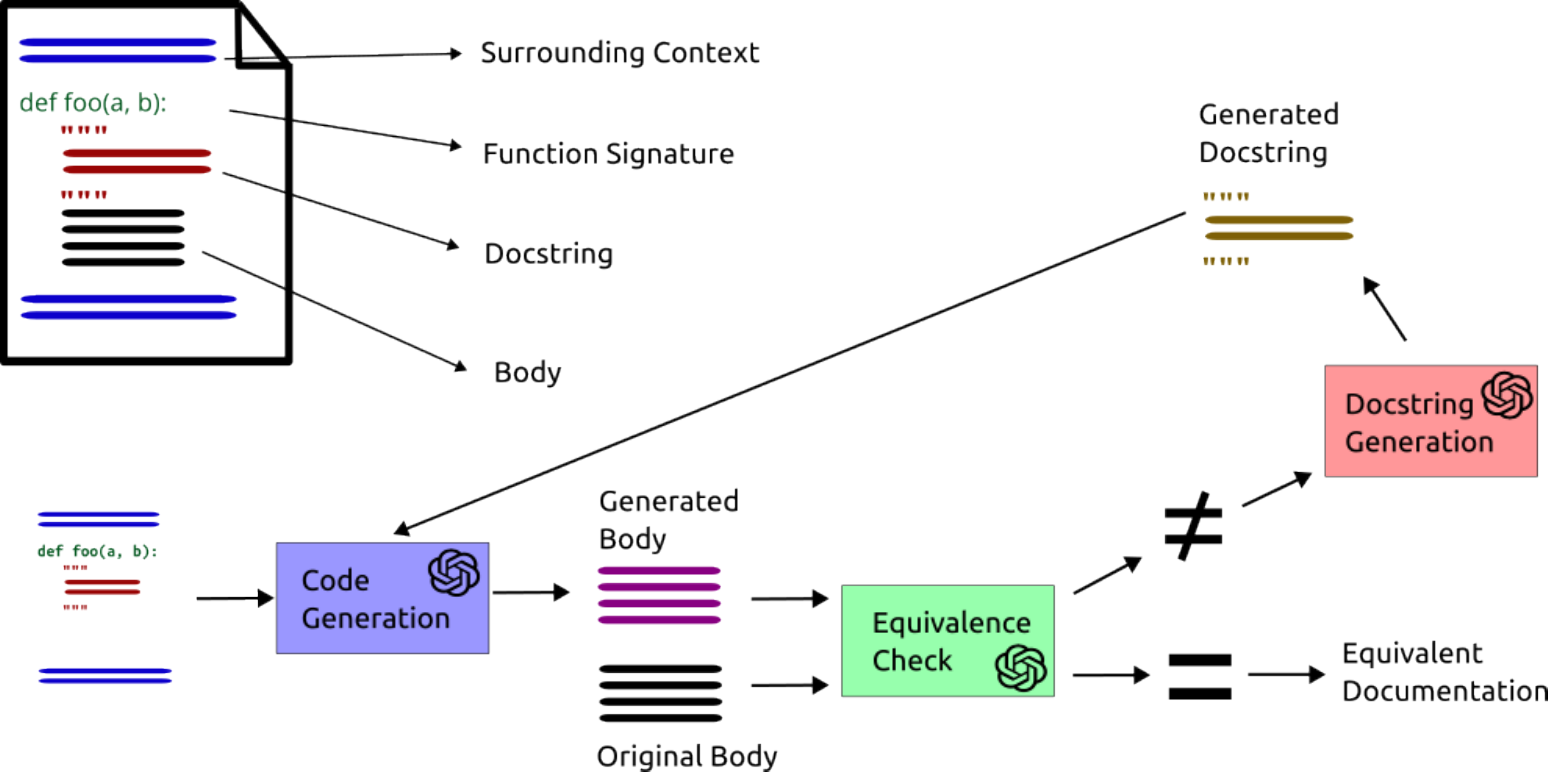}
\caption{Overview of the iterative approach to generate equivalent documentation.}
\label{fig:overview}
\end{figure}

We present an automatic approach, called \name{}, to generate equivalent documentation.
Given a function signature, its body, and the surrounding context, \name{} produces a docstring that is equivalent to the function body, according to the definition in \cref{sec:definition}.

\Cref{fig:overview} shows an overview of \name{}.
\name{} starts by generating code from the existing documentation (if none exists, it uses only the surrounding context).
Then, using the same LLM that generated the code as a judge~\cite{Ahmed2025,Bavaresco2025}, the approach compares the generated code with the original code to determine whether they are equivalent.
If they are not equivalent, the approach generates a brief description of the differences between the two code snippets.
This description is then used to generate new documentation that is more specific and accurate.
\name{} repeats this process until the generated documentation is equivalent to the given code snippet, i.e., the code generated from the documentation is judged to be equivalent to the original code, or until a maximum number of iterations is reached.

This design is motivated by two observations about the capabilities of LLMs.
First, we observe that LLMs are good at generating code and documentation, but they sometimes miss some important details or, because of missing context, their generated code or documentation may not be equivalent to the ground truth.
The approach builds on this observation by leveraging LLMs to generate code and documentation in a roundtrip manner, which allows us to iteratively refine the documentation.
Second, we observe that LLMs are good at comparing pieces of code and identifying semantic differences between them.
\name{} leverages this ability to identify the differences between code generated from candidate docstrings and the original code, and to use any differences to guide the model to generate more complete and accurate documentation.

\subsubsection{Core Algorithm}
\label{sec:main_loop}

The core of our approach is an iterative process that generates and refines a docstring for a given function.
\Cref{alg:main_loop} summarizes this process.
We describe the details of each part of the algorithm in the following sections.

The first iteration of the main loop (lines~\ref{line:main_loop_start}--\ref{line:main_loop_end}) starts with generating the body of the function (described in more detail in \cref{sec:code_generation}) from the surrounding context and the existing docstring, if any.
The reason for starting with the existing docstring is that it may already be equivalent to the code, in which case we can stop immediately.
Even if the existing docstring is not equivalent, it can serve as a basis for generating a new docstring.

\begin{algorithm}[t]
\caption{Algorithm to generate an equivalent docstring.}
\label{alg:main_loop}
\begin{algorithmic}[1]
\State \textbf{Input:} function\_header, function\_body, surrounding\_context, size\_limit, max\_iterations
\State \textbf{Output:} equivalent docstring or failure
\State docstring = original\_docstring or ""
\State iterations = 0
\State equivalent = False
\While{iterations < max\_iterations}  \Comment{Iterative refinement of docstring} \label{line:main_loop_start}
    \State generated\_body = generate\_body(surrounding\_context, function\_header, docstring)
    \State equivalent = check\_equivalence(function\_body, generated\_body)
    \If{equivalent}
        \State \textbf{return} docstring \label{line:return_equivalent1}
    \EndIf
    \State differences = generate\_differences(function\_body, generated\_body)
    \State docstring = generate\_docstring(docstring, differences)
    \State iterations += 1
\EndWhile \label{line:main_loop_end}
\If{\textbf{not} equivalent} \Comment{Final attempt to refine docstring} \label{line:final_refinement_start}
    \State docstring = refine\_docstring(docstring, function\_body) \label{line:final_refinement}
    \State generated\_body = generate\_body(surrounding\_context, function\_header, docstring) \label{line:generate_next_body} \label{line:final_equivalence_check_start}
    \State equivalent = check\_equivalence(function\_body, generated\_body) \label{line:final_equivalence_check_end}
    \If{equivalent}
        \State \textbf{return} docstring
    \EndIf
\EndIf
\State \textbf{return} failure
\end{algorithmic}
\end{algorithm}

The generated body is then compared with the original body to check if they are equivalent, which determines if we need to continue with the approach or can stop and return the docstring (line~\ref{line:return_equivalent1}).
The equivalence check is described in more detail in \cref{sec:equivalence_check}.
If the generated body is not equivalent to the original body, we generate a brief description of the differences between the two code snippets, as described in \cref{sec:docstring_generation}.
These differences help the model incorporate the missing information into the new docstring in the next step (line~\ref{line:generate_next_body}).
The new docstring is generated based on the surrounding context, the original body, the generated body, the differences between the two bodies, and the previous docstring.
This completes the main loop, with the generated docstring being used in the next iteration to generate a new body.

The main loop runs for at most a configurable number of iterations (default: five), or until an equivalent docstring is found.
If the approach reaches the maximum number of iterations but does not find an equivalent docstring (line~\ref{line:final_refinement_start}), the generated docstring may contain information that does not match the given implementation.
For example, the docstring may mention checking for a certain exception, whereas the implementation does not check for that exception.
To handle such cases, the algorithm includes a final refinement step (line~\ref{line:final_refinement}).
In this step, we prompt the LLM to remove any information from the generated docstring that does not correspond to any code in the correct implementation.
After this step, we ask the LLM to do the first two steps of the main loop (generating the body, and evaluating the equivalence) once again (lines~\ref{line:final_equivalence_check_start}--\ref{line:final_equivalence_check_end}).
If the generated body is judged to be equivalent to the correct implementation, the approach considers the docstring equivalent.
Otherwise, the approach gives up and reports a failure.

\subsubsection{Code Generation}
\label{sec:code_generation}

The first step in the main loop is generating the body from the current candidate docstring, the function signature, and the surrounding context.
To this end, \name{} prompts the LLM based on the template shown in \cref{fig:body_prompt}.
The docstring is initially the original docstring, i.e., the docstring written by the developer, if it exists, or an empty string otherwise.
In subsequent iterations, \name{} uses the docstring generated in the previous iteration.
The surrounding context includes the whole source code file, except for the code of the function being documented.

\begin{figure}
  \begin{minted}{markdown}
In the file containing
```python
{context}
```
implement the body of the function
```python
{function_header}
{docstring}
```
Only output the body, with correct indentation. Do not repeat the function signature or docstring.
  \end{minted}
  \caption{Template of the prompt for generating the function body.}
  \label{fig:body_prompt}
\end{figure}

Given the full file context, the LLM generates the body of the function \code{soft\_br}, shown in \cref{fig:running_example_generated_body}, which is different from the original body shown in \cref{fig:running_example_original}.
Note that although the generated body is different, it may still be functionally equivalent to the original body.
Therefore, we need to continue with the next step, which is the equivalence check, to determine if the generated body is indeed equivalent to the original body.

\begin{figure}
  \begin{minted}{python}
      if self.blockquote > 0:
          self.o("\n> ")
      else:
          self.o("\n")
  \end{minted}
  \caption{Generated body for the function \code{soft\_br} from the original docstring.}
  \label{fig:running_example_generated_body}
\end{figure}

\subsubsection{Equivalence Check}
\label{sec:equivalence_check}

To determine whether \name{} has achieved its goal of generating an equivalent docstring, the approach checks if the code generated from the current docstring is equivalent to the original code.
This can be done in various ways, for example by executing both code snippets on a set of inputs and comparing the outputs, by using string matching, or by employing an LLM to judge their equivalence.
String matching is too strict and can cause false negatives, as two code snippets can be equivalent while being syntactically different.
Executing the code snippets is a more robust way to check for equivalence, but it requires a set of inputs that cover the behavior of the function, which may not be available, and are hard to generate automatically.
Moreover, running an arbitrary piece of code is not always feasible~\cite{Souza2025}.
Using an LLM as a judge is a more flexible approach that can be applied to any code snippet.
It also allows the approach to be language and platform agnostic.
Therefore, \name{} compares the generated code with the original code using an LLM as judge, which assesses whether the two code snippets are equivalent.

\begin{figure}
  \begin{minted}{markdown}
Are the following two code pieces equivalent?
Correct implementation:
```python
{function_header}
{function_body}
```
Alternative implementation:
```python
{function_header}
{generated_body}
```
Briefly explain why the alternative implementation is or is not equivalent to the correct implementation, at a low level.
Phrase the differences as "The correct implementation checks for X, but the alternative implementation does not", "The correct implementation calls Y, but the alternative implementation calls Z", etc.
Only mention each difference once.
If possible, give an example case where they behave differently.
Ignore all implicit errors and exceptions.
If there are cases where they behave differently, output "DIFFERENT" at the end of the response.
Otherwise, if they behave exactly the same for all inputs, then they are equivalent, and you should output "EQUIVALENT" at the end of the response.
The output must contain one of "DIFFERENT" or "EQUIVALENT", never both.
  \end{minted}
  \caption{Template of the prompt for checking the equivalence of two code snippets.}
  \label{fig:equivalence_prompt}
\end{figure}

To compare the two code snippets, \name{} prompts the LLM with a prompt, as shown in \cref{fig:equivalence_prompt}.
The prompt contains both code snippets and asks the LLM to first describe the differences between the two snippets if there are any, and then answer whether the two snippets are equivalent or not.
The prompt also asks the LLM to provide an example of input that results in different behavior in the two code snippets, if possible.
The explanation of differences and input examples provide a chain-of-thought reasoning for the LLM, which has been shown to improve the quality of the responses~\cite{Wei2022}.

Finally, the prompt directs the model to ignore any implicit errors and exceptions.
This is because the LLM does not know about the domain of valid inputs for the function, and may consider two code snippets to be different if one of them raises an exception for an input that is not valid for the function.
To avoid such false negatives, we ask the model to ignore implicit errors and exceptions.

If the LLM decides that the two code snippets are equivalent, the process is terminated and the docstring is considered equivalent to the code.
Otherwise, we move to the next step, which is generating a new docstring.

\cref{fig:running_example_equivalence} shows the response from the LLM when comparing the generated body in \cref{fig:running_example_generated_body} with the original body in \cref{fig:running_example_original}.
It first describes three differences between the two implementations, namely, calling different functions, modifying different state, and checking for a condition that the other implementation does not check for.
Then, based on the last point of difference (the condition check), it provides an example of an input that results in different behavior in the two implementations.
Finally, it outputs \code{DIFFERENT}, indicating that the two implementations are not equivalent.
Since the verdict of \name{} is that the two implementations are not equivalent, the approach continues with the main loop, which will generate a new docstring.

\begin{figure}
  \begin{minted}{markdown}
Not equivalent.

- The correct implementation calls self.pbr(), but the alternative implementation calls 
self.o(...) to write a newline (and potentially a blockquote prefix in the alternative path).
- The correct implementation assigns self.br_toggle = \"  \", but the alternative
implementation does not modify br_toggle.
- The correct implementation does not check self.blockquote, but the alternative
implementation checks if self.blockquote > 0 to decide between \"\\n> \" and \"\\n\".

Example where they differ: If self.blockquote is greater than 0, the alternative will output 
a line starting with \"> \" after the newline, whereas the correct version does not emit 
that blockquote prefix (and may affect subsequent formatting relying on br_toggle or other 
state). This can lead to different rendered output after soft_br().

DIFFERENT
  \end{minted}
  \caption{Response from the LLM when comparing the generated body with the original body.}
  \label{fig:running_example_equivalence}
\end{figure}

\subsubsection{Docstring Generation}
\label{sec:docstring_generation}

If the generated code is not equivalent to the original code, our approach attempts to generate a new, revised docstring.
The prompt includes the surrounding context, the developer's implementation of the function, and the docstring and body generated in the previous iteration.
We also calculate the diff of the correct and generated function body, so that the LLM can focus on the differences.

\begin{figure}
  \begin{minted}{markdown}
In the file containing
```python
{context}
```
the correct implementation of the function
```python
{function_header}
```
is
```python
{function_header}
{function_body}
```
However, using the docstring
```python
{new_docstring}
```
the body was generated as
```python
{function_header}
{generated_body}
```
Here is the diff between the generated body (before) and the correct implementation (after):
```diff
{diff}
```
First, explain all the ways in which the correct implementation is different from the generated body.
Then, for each difference, identify which part of the docstring is responsible for that difference (if any) and fix that part of the docstring to represent the correct implementation.
The docstring must not mention the generated body.
If there are any differences that are not related to any part of the docstring, add a sentence to the docstring explaining the correct implementation of that part, such as custom functions, specific checks, or other details that are not in the generated body.
Do not add any new information that is not in the correct implementation.
The wording of the docstring must be as if it is describing the correct implementation without mentioning the phrases like "correct implementation" or "correct body".
The docstring must be in natural language and start and end with triple double-quotes.
Output the docstring inside triple backticks with correct indentation (`{indentation}`). Do not include the function header or body.
Limit the new docstring to a maximum of {int(size_limit * orig_function_length)} lines.
  \end{minted}
  \caption{Template of the prompt for generating a new docstring.}
  \label{fig:docstring_prompt}
\end{figure}

In this prompt, as seen in \cref{fig:docstring_prompt}, we ask the LLM to first describe the differences between the generated body and the correct body.
As in the equivalence check, the LLM is prompted to explain the differences, providing chain-of-thought reasoning.
We ask the LLM to then identify the part of the docstring responsible for the code that is different and modify the docstring to match the correct implementation.
For differences that have no correspondence with the docstring, we ask the model to add the necessary information to the docstring.
We also add instructions for the LLM to limit the length of the docstring to the provided size limit.
We specify the size limit in terms of a multiple of the original function length, which defaults to 1x, i.e., the same length as the original function.

Given the prompt described above, the model generates a new docstring.
\name{} checks whether the docstring adheres to the size limit.
If the response exceeds the size limit, we prompt the LLM again to summarize the docstring to fit in the size limit.

For the example function \code{soft\_br}, the LLM generates the docstring shown in \cref{fig:running_example_generated_docstring}.
This docstring now explains what the function does, and how it works.
In the next iteration of the main loop, this docstring is used to generate a new body, which matches the original body.
The equivalence check of \name{} also evaluates the two code pieces as equivalent, and so the docstring is returned.

\begin{figure}
  \begin{minted}{python}
"""Soft line breaks are produced by signaling a paragraph boundary and preparing a two-space continuation.
It first marks a paragraph break, then assigns br_toggle to two spaces, so the next line ends with two spaces before the newline.
This preserves Markdown-style line breaks within paragraphs."""
  \end{minted}
  \caption{Generated docstring for the function \code{soft\_br} after the first iteration.}
  \label{fig:running_example_generated_docstring}
\end{figure}

\section{Applications}

We consider two common software engineering tasks---output prediction and code editing---and in our evaluation, show how equivalent documentation benefits both human developers and LLMs when performing these tasks.

\subsection{Code Understanding via Output Prediction}
\label{sec:app_code_understanding}

The first task we consider is code understanding, where we use output prediction as a proxy task.
In this task, given a function signature and its documentation, the goal is to predict the output of the function for a given input.
This task reveals how much behavioral information the documentation contains about the function.
Since many downstream software engineering tasks, such as test generation, bug detection, and code editing, require understanding the behavior of the code on a certain input, improvements in this task can also benefit those tasks.

\begin{figure}
  \begin{minted}{python}
    def __init__(self, app: ASGIApp) -> None:
        allow_origins_str = os.getenv('PERMITTED_CORS_ORIGINS')
        if allow_origins_str:
            allow_origins = tuple(origin.strip() for origin in allow_origins_str.split(','))
        else:
            allow_origins = ()
        super().__init__(app, allow_origins=allow_origins allow_credentials=True,
            allow_methods=['*'],allow_headers=['*'],)
  \end{minted}
  \caption{Example of a function from the \code{All-Hands-AI/OpenHands} repository.}
  \label{fig:function_example}
\end{figure}

\begin{figure}
  \begin{minted}{python}
def test_localhost_cors_middleware_init_without_env_var():
    """Test that the middleware works correctly without PERMITTED_CORS_ORIGINS environment variable."""
    with patch.dict(os.environ, {}, clear=True):
        app = FastAPI()
        middleware = LocalhostCORSMiddleware(app)
        # Check that allow_origins is empty when no environment variable is set
        assert middleware.allow_origins == 
  \end{minted}
  \caption{Example of a test prefix that exercises the function in \cref{fig:function_example}. The correct value for the incomplete assertion is \code{()}, i.e., an empty tuple.}
  \label{fig:test_example}
\end{figure}

More formally, we define the task as follows.
Given the signature of a function under test $f$, its documentation $d$, and a test prefix $p$ that exercises $f$ and ends with an assertion statement with one missing value $v$, the goal is to predict $v$.
An example of such a function is shown in \cref{fig:function_example}, which is the constructor of a middleware class that reads an environment variable to set the allowed origins for CORS requests.
An example of a test prefix that exercises this function is shown in \cref{fig:test_example}, which tests the behavior of the function when the environment variable is not set.
The test prefix ends with an assertion statement that checks the value of the \code{allow\_origins} attribute, but the expected value is missing.
The goal is to predict the missing value, which in this case is an empty tuple, or \code{()} in Python.

\subsection{Code Editing}
\label{sec:app_code_change}

As the second task, we consider code editing.
This task represents a scenario in which a developer wants to modify a function based on a natural language description of the change.
In this task, given a code snippet $c_{old}$ with its documentation $d_{old}$ and a change description $m$, the goal is to modify the code to $c_{new}$.

For example, consider commit \code{c718248} in the \code{keras-team/keras} repository, which describes the change in the code as "Validate positive height and width in image resize".
\Cref{fig:keras_code_change} shows the change in function \code{resize}, where a check for negative dimensions is added that raises a \code{ValueError} if either dimension is negative.

\begin{figure}
  \centering
  \begin{minted}{diff}
    if len(size) != 2:
        raise ValueError(
            "Expected `size` to be a tuple of 2 integers. "
            f"Received: size={size}"
        )
+   if size[0] <= 0 or size[1] <= 0:
+       raise ValueError(
+           f"`size` must have positive height and width. Received: size={size}"
+       )
    if len(images.shape) < 3 or len(images.shape) > 4:
        raise ValueError(
            "Invalid images rank: expected rank 3 (single image) "
  \end{minted}
  \caption{Example of a code change from the \code{keras-team/keras} repository.}
  \label{fig:keras_code_change}
\end{figure}

\section{Evaluation}
We evaluate our approach and the utility of the generated documentation in downstream tasks, addressing the following research questions:
\begin{enumerate}[label=RQ\arabic*]
    \item How effective is our approach in generating equivalent docstrings?
    \item How transferable are equivalent docstrings across different LLMs?
    \item How do different components and parameters of the approach affect the generated docstrings?
    \item What computational costs does the approach incur?
    \item How much do equivalent docstrings improve LLM performance on the code understanding task?
    \item How much do equivalent docstrings improve human performance on the natural language-based code editing task?
\end{enumerate}

\subsection{Experimental Setup}
\subsubsection{Datasets}
\label{sec:datasets}
We use two datasets from previous work as the basis for our evaluations.
The first dataset is CoDocBench~\cite{Pai2025}, which is a dataset of pairs of code and docstring changes in the history of 203 popular Python projects on GitHub.
The second dataset is DyPyBench~\cite{fse2024-DyPyBench}, which is a dataset of 50 popular projects across various domains that have executable tests.
From the union of these datasets (252 unique projects), we create three datasets for our evaluation.

The first dataset, which we call the \emph{code-docstring equivalence dataset}, is used for RQs1--4, where we want to evaluate the behavioral equivalence of the generated code piece with the human-written code.
Therefore, we use test execution as the ground truth for behavioral equivalence.
To this end, for each project in the union of the datasets, we try to run the test suite by installing the dependencies using the \code{pipreqs} package\footnote{\url{https://github.com/bndr/pipreqs}}.
From the executable projects, we take the functions that are covered at least 50\% by the tests.
This results in 2437 functions from 12 projects.
We then sample at most 25 functions from each project, which results in a total of 238 functions.
These functions range from 2 to 168 lines of code, with an average of 16 and median of 9 lines of code.
Most functions lack docstrings; only 47\% are documented.

For RQ5, we use the same set of 252 projects described above.
Then we extract the tests that have assertions with a comparison, where one side of the comparison is either a literal or an explicit collection, for example \code{assert foo() == 42} or \code{assertEqual(foo(), [1, 2, 3])}.
We then use the heuristic by \citet{Watson2020} to find the function-under-test for each test.
We exclude test cases where \code{gpt-4.1-nano} is not able to predict all the assertion values correctly given the implementation of the function under test, as we consider these cases too hard for the LLM to predict.
This results in 253 test functions, with a total of 480 assertions, which we call the \emph{output prediction dataset}.

Finally, we compile a third dataset, which we call the \emph{user study dataset} for RQ6.
From five popular projects, namely, \code{keras}, \code{marshmallow}, \code{celery}, \code{requests}, and \code{click}, we select the latest commit (from \code{keras} the last two commits) where the code change is isolated to one function with fewer than 10 lines of change, the change is not trivial from the commit message, and there is a test that exercises the changed function.
This process ensures that there is a need for understanding the code to perform the change, that we can use test execution to check the correctness of the change, and that the participants are less likely to have seen the change before.

\subsubsection{Baseline}
\label{sec:baseline}
We implement a baseline for our evaluations, which is prompting the LLM to generate a docstring for the given function in a single step.
We use the same LLM as in \name{}.
The prompt, as shown in \cref{fig:baseline}, includes the surrounding context, the function signature, and the function body.

\begin{figure}
  \begin{minted}{markdown}
In the file containing
```
{context}
```
generate a Python docstring for function:
```
{function_code}
```
Only output the docstring with correct indentation. Do not repeat the function signature or the function body.
  \end{minted}
  \caption{The template of the prompt for the baseline approach.}
  \label{fig:baseline}
\end{figure}

\subsubsection{LLMs and Hardware}
We run all experiments on an Ubuntu 22.04 machine with Intel Xeon CPU with 48 cores running at 2.20GHz and 256GB of RAM.
The experiments are run inside Python 3.12 Docker containers.
We use the \code{gpt-4.1-nano-2025-04-14} as the main LLM through the OpenAI API.

\subsection{RQ1: Generating Equivalent Documentation}
\label{sec:eval_equivalence}
We run \name{} on each function in the dataset to obtain an equivalent docstring.
For cases where \name{} fails to generate an equivalent docstring, we use the last generated docstring as its best effort.
Then we prompt the LLM to generate the body of the function, based on the context, the function signature, and the generated docstring.
Finally, we run the tests to check if the generated code behaves the same as the original code.
If the test results of the generated code match the test results from the original code, we consider the generated docstring to be equivalent to the code.
We report the success rate, and precision, which show how often \name{} generates a docstring equivalent to the code, and how often docstrings that \name{} claims to be equivalent are actually equivalent, respectively.

As shown in \cref{tab:rq1}, \name{} generates equivalent docstrings in 53.4\% of the cases.
When claiming that a generated docstring is equivalent to the code, then for 66.4\% of these cases the generated docstrings are indeed equivalent to the code.

\begin{figure}
  \centering
  \includegraphics[width=0.8\textwidth]{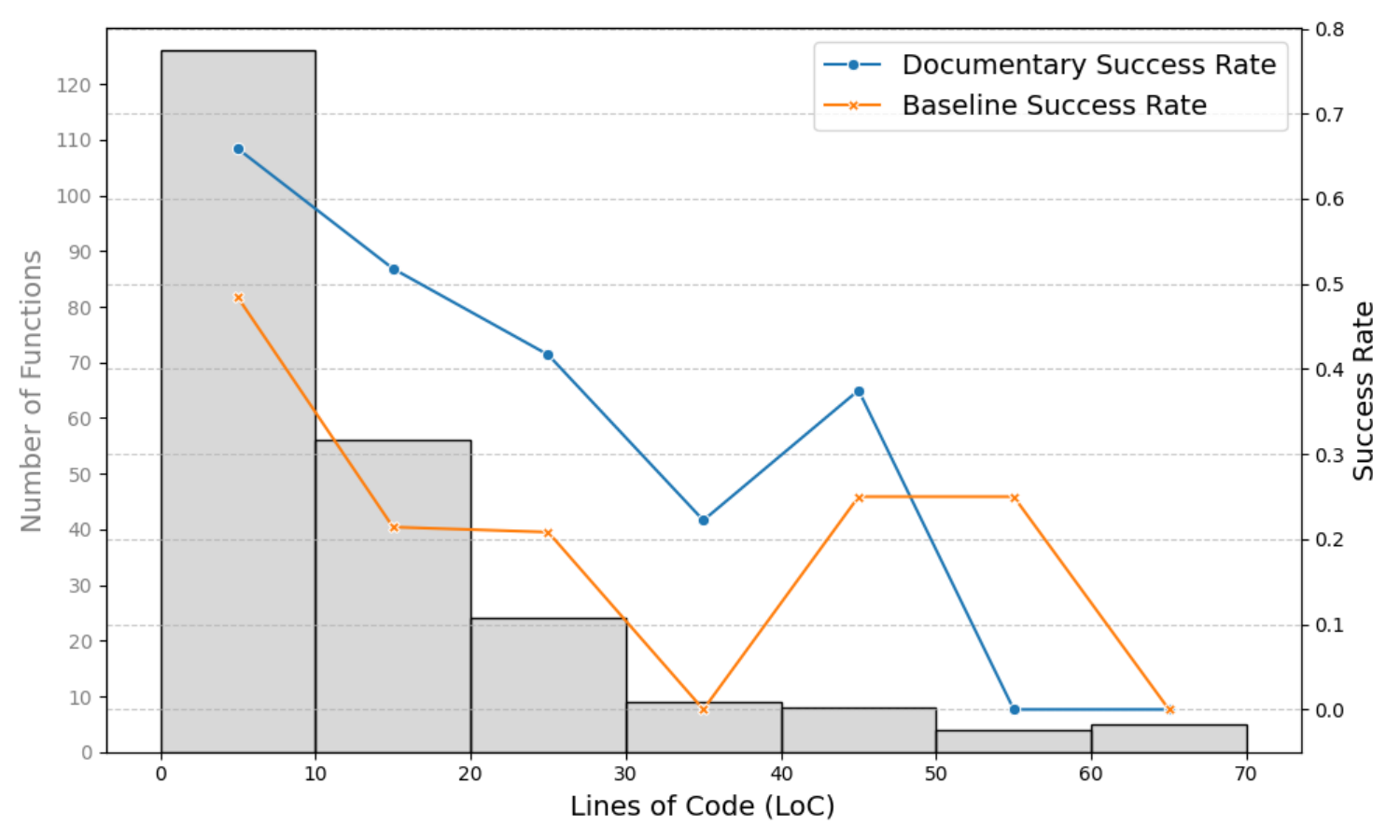}
  \caption{Distribution of successful generation of equivalent documentation using \name{} and the baseline for different function lengths.}
  \label{fig:loc_distribution}
\end{figure}

\Cref{fig:loc_distribution} shows how the length of the function, measured in lines of code (LoC), affects the success rate of generating equivalent documentation.
Unsurprisingly, the success rate decreases as the function length increases, which is expected as longer functions are more complex and harder to describe in a docstring.
However, \name{} consistently outperforms the baseline across all function lengths except one, where there is only one function that the baseline is able to generate equivalent docstring.

\begin{table}
  \caption{Effectiveness in generating equivalent documentation. Success shows how often the approach generates an equivalent documentation based on test execution. Total is the number of functions in the dataset, and Precision shows the ratio of true positives to all cases that \name{} claims to have generated equivalent documentation.}
  \label{tab:rq1}
  \centering
  \begin{tabular}{lrr}
    \hline
    Approach & Success / Total (\%) & Precision \\
    \hline
    Baseline & 81 / 238 (34.0\%) & 34.0\% \\
    \name{} & \textbf{127} / 238 (\textbf{53.4\%}) & \textbf{66.4\%} \\
    \hline
  \end{tabular}
\end{table}

\subsection{RQ2: Model Independence}
To evaluate how useful equivalent documentation is across different models, we run an experiment where we generate equivalent documentation using one LLM, and then use that documentation to generate code using different LLMs.
We then run the tests to check if the generated code is equivalent to the original code.
If the test results match the test results from the original code, we consider the generated code to be equivalent to the original code.
For the model used in \name{}, we use \code{gpt-4.1-nano}, and we use \code{gpt-5-nano}, and \code{gemini-2.5-flash-lite} as the other LLMs used for generating code from the equivalent documentation.

For each function, we first generate equivalent documentation using \code{gpt-4.1-nano} and our approach.
Then, for cases where \name{} declares the generated documentation as equivalent, we use the generated documentation to generate code using \code{gpt-5-nano}, and \code{gemini-2.5-flash-\allowbreak lite}.
Finally, we run the tests to check if the generated code is equivalent to the original code.
For \code{gpt-5-nano}, 89.6\% of the generated code are equivalent to the original code, and for \code{gemini-2.5-\allowbreak flash-lite}, 74.7\% of the generated code are equivalent to the original code.
Comparing these results with \name{}'s precision of 88.9\% (obtained when using \code{gpt-4.1-nano} for code generation), we see that other LLMs generate equivalent code at a similar rate.

These results show that equivalent documentation can be generated and used by different LLMs.
More importantly, this also shows that equivalent documentation can persist in the code and provide its benefits while the LLMs evolve.

\subsection{RQ3: Ablation Study}
\begin{figure}
\centering
\includegraphics[width=0.7\textwidth]{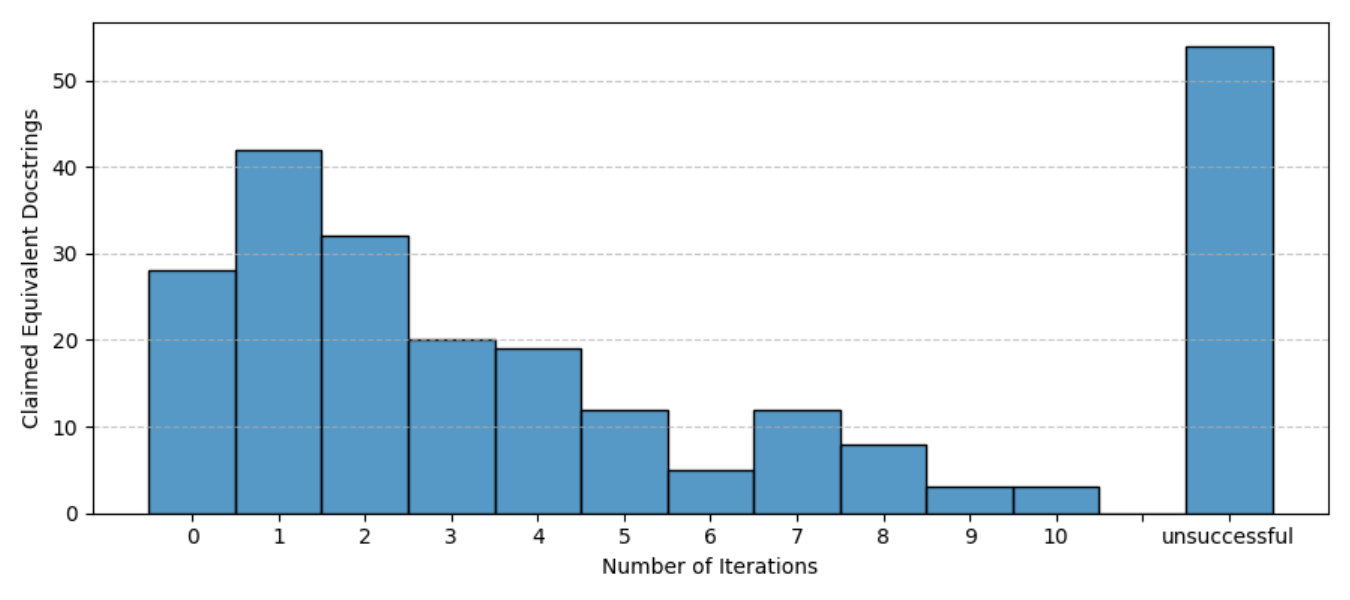}
\caption{Distribution of the number of iterations when generating equivalent docstrings.}
\label{fig:ablation_iterations}
\end{figure}

We study the effects of the number of iterations and the size limit on the effectiveness of generating equivalent documentation.
For the number of iterations, we run our approach with a maximum of 10 iterations and a size limit of 2x, and report the distribution of the iteration count when the approach decides that the generated documentation is equivalent to the code.
If the original docstring is already equivalent to the code, the iteration count is zero.
Iteration counts greater than zero mean that the approach generates the docstring after that many attempts.
As shown in \cref{fig:ablation_iterations}, most (75\%) equivalent documentation is generated within five iterations, and the median number of iterations is two.

\begin{figure}
\centering
\includegraphics[width=0.7\textwidth]{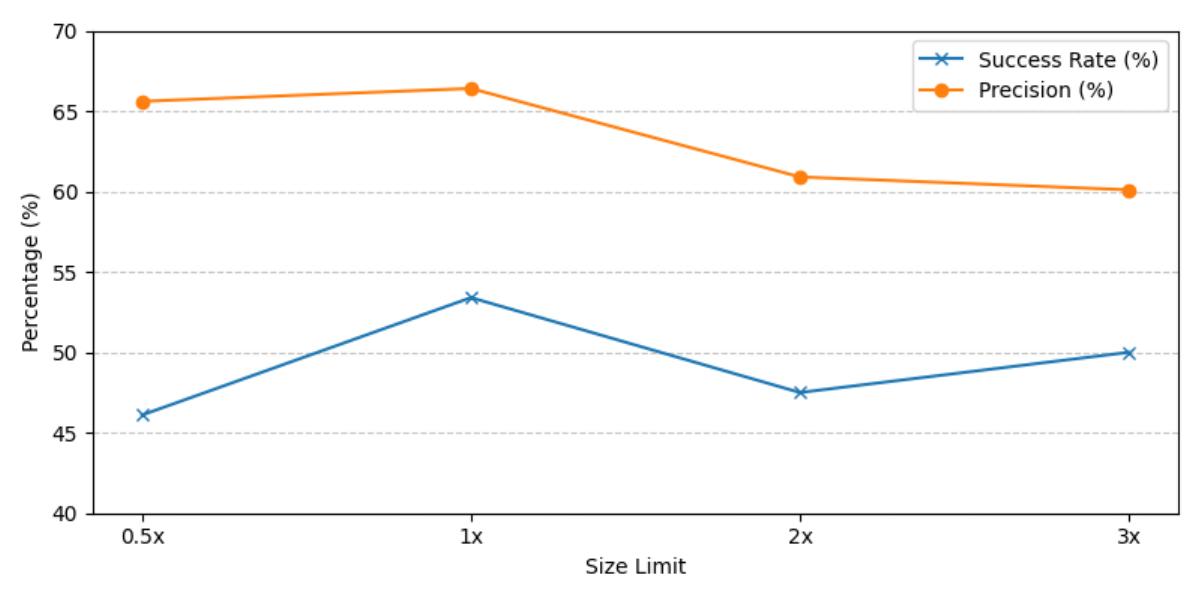}
\caption{Success rate and precision of generating equivalent docstrings for different size limits.}
\label{fig:ablation_size_limit}
\end{figure}

For the size limit, we run our approach with a maximum of 10 iterations and size limits of 0.5x, 1x, 2x, and 3x.
\Cref{fig:ablation_size_limit} shows the success rate and the precision of the generated equivalent documentation for different size limits.
Our results show that a size limit of 1x provides the best precision and success rate.
We suspect that a size limit greater than 1x encourages the LLM to generate verbose docstrings that contain unnecessary information, which can confuse the LLM in generating the code from the docstring.

\subsection{RQ4: Costs}
We measure the runtime and monetary costs of generating equivalent documentation using \name{}.
This experiment measures the time it takes to run \name{} per function in seconds using Python's \code{time.perf\_counter}, as well as the number of input and output tokens when running \name{} on a function.
We then calculate the cost of generating an equivalent docstring, based on the current pricing of \code{gpt-4.1-nano}.

\begin{figure}
\centering
\includegraphics[width=\textwidth]{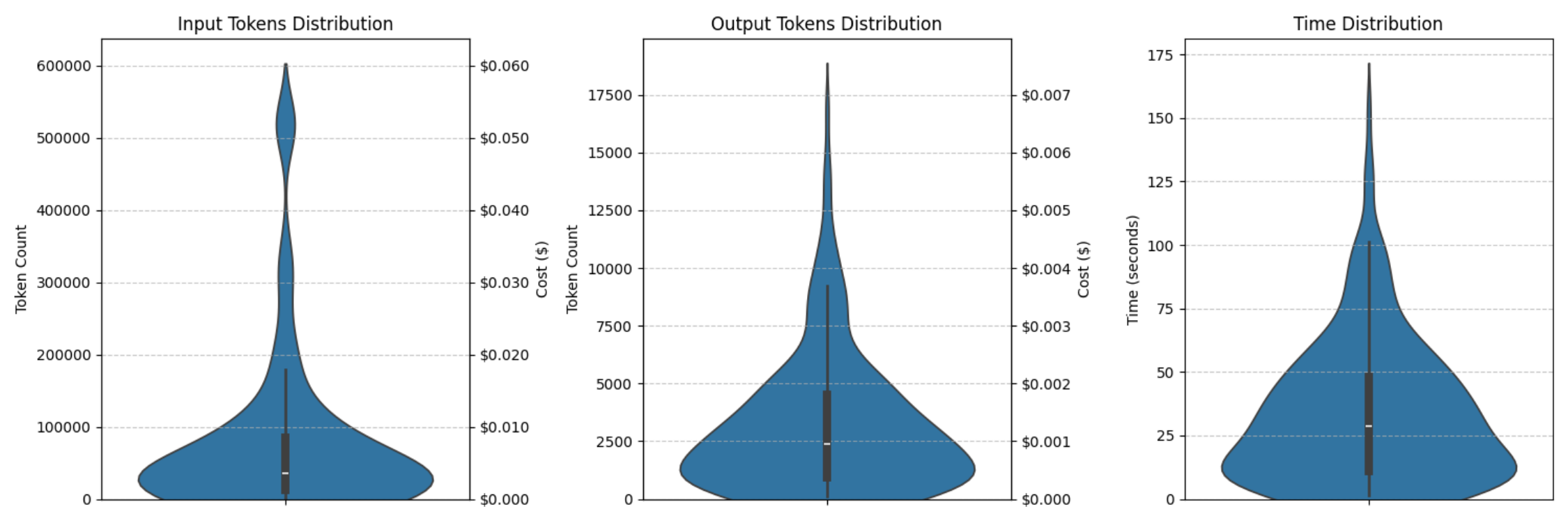}
\caption{Distributions of the number of input and output tokens, and the runtime of generating equivalent docstrings.}
\label{fig:costs}
\end{figure}

Regarding time, it takes 33.8 seconds to generate an equivalent docstring on average per function.
\Cref{fig:costs} shows that for most functions it takes less than a minute to generate an equivalent docstring.
Regarding token consumption, \name{} uses 78,144 input tokens and 3,173 output tokens, on average, for an attempt at generating an equivalent docstring for a function.
Given the pricing of \code{gpt-4.1-nano} at \$0.10 per 1M input tokens and \$0.40 per 1M output tokens in February 2026, the average cost of generating an equivalent docstring for a function is \$0.009.
As shown in \cref{fig:costs}, most functions require less than 200K input tokens and 10K output tokens, resulting in a cost of less than \$0.03 per function.

Furthermore, for large codebases to adopt equivalent documentation, the cost can be reduced by batching the generation and using cached input tokens as the context for multiple functions that contain the same prefix.
Given the reasonable costs of our current approach, we leave such optimizations for future work.

\subsection{RQ5: Equivalent Documentation for LLMs (Code Understanding)}
We run an experiment to evaluate how well equivalent docstrings help LLMs understand the behavior of a function without seeing the code, as specified in \cref{sec:app_code_understanding}.
For this experiment, we use the \emph{output prediction dataset} described in \cref{sec:datasets}.
For each test in the dataset, we first generate an equivalent docstring for the function-under-test using \name{}.
If \name{} fails to generate an equivalent docstring, we fall back to using a docstring generated by the LLM, as in the baseline.
Then, we prompt the LLM with the function signature, the equivalent docstring, and the test code, and ask it to predict the output of each assertion in the test as a valid Python expression.
Finally, we compare the AST of the predicted output with the AST of the actual output in the assertion statement, and consider a match as a correct prediction.

The metrics we report are the number of correctly predicted assertions, the average percentage of correctly predicted assertions per test, and the number of tests where all assertions are correctly predicted.
We compare the results of \name{} with the predictions from the original docstring and the baseline defined in \cref{sec:baseline}.

\Cref{tab:rq5} shows that using equivalent docstrings improves the number of correctly predicted assertions by 4.0--15.9, the average percentage of correctly predicted assertions per test by 0.6--17.3, and the number of tests where all assertions are correctly predicted by 5.6--19.0 absolute percentage points, compared to using human-written docstrings, and LLM generated docstrings.
The results show that an equivalent docstring allows the LLM to not only implement the correct behavior, as shown in \cref{sec:eval_equivalence}, but also predict the behavior of the function.

\begin{table}
  \caption{Effectiveness in output prediction. Correct Assertions shows how many assertions are correctly predicted, and Correct Tests shows how many tests have all assertions correctly predicted.}
  \label{tab:rq5}
  \centering
  \begin{tabular}{lrr@{\qquad}r@{\qquad}rr}
    \toprule
     & \multicolumn{2}{l}{Correct Assertions} & \multicolumn{1}{c}{ \multirow{2}{9em}{Average Correct Assertions per Test} } & \multicolumn{2}{c}{Correct Tests} \\ 
    \cmidrule(r{12pt}){2-3} \cmidrule{5-6}
    Approach & Count & \% & & Count & \% \\
    \midrule
    Original docstring & 280 / 480 & 58.3\% & 53.5\% & 105 / 253 & 41.5\% \\
    Baseline & 337 / 480 & 70.2\% & 70.2\% & 139 / 253 & 54.9\% \\
    \name{} & \textbf{356 / 480} & \textbf{74.2\%} & \textbf{70.8\%} & \textbf{153 / 253} & \textbf{60.5\%} \\
    \bottomrule
  \end{tabular}
\end{table}

For the example in \cref{fig:function_example}, \name{} generates the equivalent docstring shown in \cref{fig:generated_docstring_rq5}.
This docstring provides a detailed explanation of the function's behavior, including how it initializes the \code{allow\_origins}, which results in a correct prediction of the assertion value, which is an empty tuple.
The LLM, however, predicts an empty list as the assertion value for both the original docstring and the baseline generated docstring.

\begin{figure}
  \begin{minted}{python}
"""
This initialization method retrieves the list of permitted CORS origins from the environment variable 'PERMITTED_CORS_ORIGINS'. 
If this variable is set, it splits the string by commas and strips whitespace from each origin, converting it into a tuple.
If the variable is not set, it defaults to an empty tuple, allowing all origins.
super().__init__ is called with the application instance and the list of allowed origins, along with standard CORS settings:
allow_credentials is enabled, and all methods and headers are permitted.
Additional configuration, such as environment-based origin control, is handled within this setup.
"""
  \end{minted}
  \caption{The equivalent docstring generated by \name{} for the function \code{\_\_init\_\_} in \cref{fig:function_example}.}
  \label{fig:generated_docstring_rq5}
\end{figure}

\subsection{RQ6: Equivalent Documentation for Human Developers (Code Understanding and Code Editing)}
We evaluate the effectiveness of equivalent docstrings for human developers by conducting a user study with 11 participants.
The demographic statistics of the participants are shown in \cref{tab:user_study_participants}.

\begin{table}
  \caption{Demographic statistics of the participants in the user study.}
  \label{tab:user_study_participants}
  \centering
  \begin{subtable}{\linewidth}
    \centering
    \caption{Numerical attributes.}
    \begin{tabular}{lrrrr}
      \toprule
      Attribute & Min & Max & Mean & Median \\
      \midrule
      Age & 23 & 31 & 26 & 25 \\
      Years of Python experience & 0 & 6 & 3.18 & 4 \\
      \bottomrule
    \end{tabular}
  \end{subtable}

  \vspace{.7em}
  \begin{subtable}{\linewidth}
    \centering
    \caption{Categorical, personal attributes (countries of the participants are anonymized to maintain their anonymity).}
    \begin{tabular}{lll}
      \toprule
      Education & Gender & Country \\
      \midrule
      Bachelor's (27.3\%) & Male (72.7\%) & Country 1 (45.5\%) \\
      Master's (63.6\%) & Female (18.2\%) & Country 2 (45.5\%) \\
      PhD (9.1\%) & Prefer not to say (9.1\%) & Country 3 (9.1\%) \\
      \bottomrule
    \end{tabular}
  \end{subtable}

  \vspace{.7em}
  \begin{subtable}{\linewidth}
    \centering
    \caption{Categorical, professional attributes.}
    \begin{tabular}{lll}
      \toprule
      AI-tools experience & Code understanding in daily work & Code editing in daily work \\
      \midrule
      Regularly (54.5\%) & Almost every day (9.1\%) & Almost every day (27.3\%) \\
      Occasionally (27.3\%) & Once/twice a week (81.8\%) & Once/twice a week (63.6\%) \\
      Rarely (9.1\%) & Fewer than once a week (9.1\%) & Fewer than once a week (9.1\%) \\
      Only a few times (9.1\%) &  &  \\
      \bottomrule
    \end{tabular}
  \end{subtable}
\end{table}

The study consists of six tasks, which we ask the participants to perform in a web-based interface.
We give a maximum of ten minutes per task.
Each task starts with showing a Python file to the participant, with a target function in that file.
The participants are free to read any part of the file, including the target function, its docstring if it exists, and any other functions or classes in the file.
Once they are satisfied with understanding the function, the participants are asked to answer the following question, to be answered on a Likert scale from 1 to 5:
\begin{itemize}
  \item How much did the docstring help you in understanding the code?
\end{itemize}

Next, the participants proceed to a new page that shows the same file and the same target function and, in addition, now also shows a natural language description that specifies a change to the code of the target function.
We directly take the relevant part of the commit message as the change description, which is written by the developers of the code change.
The participants are asked to modify the code of the target function according to the change description.
Once they are done with the change, they submit the modified code.
The study interface then shows them three further questions about the task, each to be answered on a Likert scale from 1 to 5:
\begin{itemize}
  \item How much did the docstring help you in modifying the code?
  \item How difficult was this task?
  \item How confident are you in your solution?
\end{itemize}
If at any point during a task the ten minute timer runs out, the participant is automatically moved to the post-task questions, and the modified code is submitted as it is at that moment.

We use a within-subject, randomized design, where each participant performs three tasks with the original docstring (control), and three tasks with the docstring generated by \name{} (treatment).
The decision of treatment-first or control-first is randomized for each participant.

In addition to the questions, we also measure the time it takes the participants from seeing the code for the first time until deciding to move to the change page, and the time it takes from seeing the change description until submitting the modified code.
We perform the Mann-Whitney U test for all metrics to check if the differences in the metrics between the control and treatment groups are statistically significant.

\begin{table}
  \caption{The average values for the Likert scale questions, the time taken for understanding and changing the code, and the total time. ↑ means higher is better; ↓ means lower is better. Bold values indicate statistically significant differences between the control and treatment groups.}
  \label{tab:user_study_results}
  \centering
  \begin{tabular}{lrrr}
    \toprule
    Metric & Original & \name{} & p-value \\
    \midrule
    Help in understanding ↑ & 2.58 & \textbf{3.76} & \textbf{0.0004} \\
    Help in modifying ↑ & 2.15 & \textbf{2.85} & \textbf{0.0425} \\
    Confidence ↑ & 2.73 & 3.30 & 0.0698 \\
    Difficulty ↓ & 3.21 & 2.82 & 0.2057 \\
    Understanding time (min) ↓ & 1.94 & 2.18 & 0.6535 \\
    Modifying time (min) ↓ & 4.53 & 4.30 & 0.7388 \\
    Total time (min) ↓ & 6.48 & 6.48 & 1.000 \\
    \bottomrule
    \end{tabular}
  \end{table}

Our results show that participants found the equivalent docstrings generated by \name{} to be more helpful in both understanding and modifying the code.
These results show that developers not only find the equivalent docstrings useful, but also prefer them over the original docstrings.
Although the participants were more confident in their tasks and found the tasks easier when using the equivalent docstrings, these differences were not statistically significant.

Considering that the tasks in the user study are from complex projects, and that the participants were not familiar with the code, we envision equivalent documentation to be helpful for contributors to open source projects, who frequently work in unfamiliar codebases.

\section{Related Work}

\subsection{Improving Documentation}
There have been multiple studies on improving the consistency of documentation with respect to the code~\cite{Liu2020_JIT,Liu2021_JIT,Panthaplackel2020a}, but their focus is on resolving inconsistencies after code changes. In contrast, our work aims to generate equivalent documentation, even if the existing documentation is consistent with the code.
Another recent work~\cite{Shi2025a} has proposed to add natural language outlines to code, which are natural language descriptions that explain smaller code pieces inside functions or classes.
These outlines have been shown to assist developers in understanding the code and performing some tasks like code changes.
Although these outlines can be regarded as another type of equivalent documentation, we do not have empirical evidence of the equivalence of these outlines to the code.

\subsection{LLM Evaluation via Self-consistency}
Recent work~\cite{Allamanis2024,Min2023} propose evaluating LLMs via self-consistency in a roundtrip sequence, where the LLM is prompted to generate documentation from code, and then the documentation is used to generate code again. The generated code is then compared to the original code. That approach is similar to our evaluation of equivalent documentation, but we focus on the generation of documentation that is equivalent to the code, rather than just evaluating the LLM's ability to generate both modalities in a roundtrip correctly.

\subsection{LLM-based Software Engineering}
LLMs and LLM-based agents have shown promise in various software engineering tasks~\cite{cacm2025-agents}, such as code comprehension~\cite{Nam2024}, code generation~\cite{Chen2025}, and program repair~\cite{Jin2023,Bouzenia2025a}.
Recent studies have also explored the use of LLMs and agentic approaches for automating even more software development processes, such as executing arbitrary code~\cite{Souza2025,Bouzenia2025}.
Our work builds towards enhancing the usability of these approaches by generating equivalent documentation that can be used by LLMs to understand and modify code.

\section{Threats to Validity}
We acknowledge the following threats to the validity of our findings.
First, our evaluation relies on the test cases to determine the equivalence of code snippets.
If the test cases are not comprehensive, they may miss important aspects of the code's behavior.
Furthermore, flaky tests may lead to incorrect conclusions about the equivalence of code snippets.
We try to mitigate this threat, which is a difficult task and an ongoing research challenge~\cite{Bouzenia2025,Souza2025}, by using large datasets for our evaluations.
Second, our evaluation focuses on Python code and docstrings.
The effectiveness of our approach may vary for other programming languages or documentation styles.
However, the principles of our approach are language-agnostic and can be applied to other languages.
Third, for our user study, we were only able to recruit 11 participants, which did not allow us to observe statistically significant differences for some metrics.

\section{Conclusion}
In this paper, we present a novel property, called documentation-to-code equivalence, in which the documentation of a code snippet is considered equivalent if it contains sufficient information to generate the same code snippet.
Equivalent documentation helps LLMs perform software engineering tasks that require understanding the code, such as output prediction.
It also allows developers to understand code better.
We propose an iterative approach, called \name{}, to generate equivalent documentation.
Our evaluation on two datasets of Python projects shows that \name{} can generate equivalent documentation for 53.4\% of functions, and that using equivalent documentation improves code comprehension for both human developers and LLMs.
Furthermore, we show that equivalent documentation can be used by LLMs other than the one used to generate them, and that the cost of generating equivalent documentation is low enough to be practical for large codebases.

\section{Data Availability}
Our code, datasets, and scripts to reproduce the results are available at \url{https://github.com/sola-st/Documentary}, and as a stable snapshot also at \url{https://doi.org/10.5281/zenodo.19514143}.

\begin{acks}
This work was supported by the German Research Foundation (DFG; projects 492507603, 516334526, and 526259073). 
\end{acks}

\bibliographystyle{ACM-Reference-Format}
\bibliography{references,referencesMichael}
\end{document}